%% file: ms.tex
\documentclass[12pt,preprint]{aastex}

\newcommand{\teff}{T$_{\rm eff}$} 
\newcommand{\hipparcos}{{\scshape{Hipparcos}} }

\begin{document}

\title{Finding cool subdwarfs using a $V-J$ reduced proper-motion 
diagram: Stellar parameters for 91 candidates}

\author{David Yong and David L. Lambert}

\affil{Department of Astronomy, University of Texas, Austin, TX 78712}
\email{tofu,dll@astro.as.utexas.edu}

\begin{abstract}

We present the results of a search for cool subdwarfs for which our 
candidates were drawn from a $V-J$ reduced proper-motion diagram constructed
by \citet{salim02}.  
Kinematic (U, V, and W) and self-consistent 
stellar parameters (\teff, log g, [Fe/H], and $\xi_t$) are 
derived for 91 candidate subdwarfs based on high resolution 
spectra.  The observed stars span 3900K $<$ \teff~$<$ 6200K and
$-2.63<$ [Fe/H] $<$ 0.25 including only 3 giants 
(log g $<$ 4.0).  Of the sample, 77 stars have MgH
lines present in their spectra.  With more than 56\% of our candidate
subdwarfs having [Fe/H]$\le-1.5$, we show that
the $V-J$ reduced proper-motion diagram readily identifies 
metal-poor stars.

\end{abstract}


\section{Introduction}
\label{sec:intro}

The chemical history of our Galaxy has been the target of numerous
theoretical and observational studies for many decades.  Much
progress has taken place where the large and 
increasing body of observed stellar abundances 
(e.g., \citealt{bdp93,mcwilliam95,ryan96,bdp03}) is complemented
by the ongoing refinement of Galactic chemical evolution models 
(e.g. \citealt{timmes95,goswami00,alc01}).  
The ultimate objective of
Galactic chemical evolution studies is to be able to 
predict accurately elemental and isotopic abundances
for any given location and time.
The spectra of cool stars show molecular bands
from which measurements of isotopic abundance ratios can be made 
(e.g., Mg from MgH, Ti from TiO).  These isotopic ratios offer a unique 
opportunity to test models of Galactic chemical evolution since 
different stellar sites and nucleosynthetic processes may be 
responsible for production of the individual isotopes.  
Comparing the observed and predicted
evolution of isotopic abundances will test aspects of the
contributions of stellar nucleosynthesis to Galactic chemical evolution
in a way that is just not possible with elemental abundances.
In order to measure isotopic ratios from molecular lines, cool subdwarfs 
(unevolved metal-poor stars that are located below the solar-metallicity 
main sequence in color-magnitude diagrams) must first be found.  

Previous studies such as \citet{ryan89} and \citet{carney94} drew candidate
subdwarfs from proper-motion catalogs and were successful in identifying
large numbers of metal-poor stars.  Subsequent detailed abundance analyses
were carried out on these targets.  Both studies 
focused upon warmer targets, neglecting
cooler subdwarfs.  Objective prism surveys which identify metal-poor stars 
by the weakening of Ca {\sc ii} H and K have also been successful in identifying
considerable numbers of metal-poor stars (e.g., \citealt{bond70,beers92,christlieb00}).
Without a reliable temperature estimate, a cool metal-poor star will have
Ca {\sc ii} lines comparable to a warmer metal-rich star 
thereby biasing the sample towards warmer stars.  
Numerous stars of spectral type M and later have recently been identified
(e.g., \citealt{delfosse99}, \citealt{hawley02},
\citealt{reid02}).  However, the overwhelming majority of these cool stars are
solar metallicity.  Only a handful of metal-poor stars with spectral type M or later
have been found (e.g., \citealt{gizis97}, \citealt{scholz99}, \citealt{lepine03}, 
and \citealt{burgasser03}).  Our target stars are cooler than those found by 
\citet{ryan89} and warmer than those found by \citet{gizis97}.

In Paper I \citep{search}, we presented the results from our first search
for cool subdwarfs.  From these subdwarfs we have measured the 
evolution of the Mg isotopic ratios, extending the \citet{gl2000}
measurements to [Fe/H]=$-2.5$ (Yong \& Lambert in prep).  
In this paper, we present the results from
our second search for cool subdwarfs, self-consistent stellar (\teff, log g, 
[Fe/H], and $\xi_t$) and kinematic (U, V, W) parameters 
for a further 91 candidates.

\section{Selection criteria}
\label{sec:criteria}

In Paper I, our goal was to identify subdwarfs with 
\teff~$<{\rm 4700K}$ and ${\rm [Fe/H]}<-1.5$, that is, 
metal-poor stars with MgH molecular features.  Following
\citet{ryan89}, we drew candidates from the New Luyten Two-Tenths (NLTT)
catalog of stars with annual proper-notions in excess of 
$0\farcs18$/year \citep{luyten79,luyten80}.  We used 
the reduced proper-motion criterion to ensure that the selected 
stars had large transverse velocities relative to the 
local standard of rest.  Since halo stars are on plunging orbits, they
are over-represented in proper-motion catalogs.
Hertzsprung first introduced the concept of the reduced proper-motion,
defined as $H=V+5\log\mu+5$,
where $V$ is the apparent magnitude and 
$\mu$ is the proper-motion in arcsec/year.  The reduced proper-motion
can also be expressed as $H=M_V+5\log\nu_T-3.37$ where $M_V$ is
the absolute magnitude and $\nu_T$ is the transverse velocity in
km/s.  At a given spectral type, a reduced proper-motion constraint
rejects stars whose transverse velocities fall below a chosen
value.  Subdwarfs are less luminous than disk dwarfs at 
a given color and have larger transverse velocities.  Both of these
factors lead to an increase in the reduced 
proper-motion of the subdwarf population
relative to the disk dwarf population.  While our original search, Paper I, was
successful in finding these subdwarfs, our efficiency was less than 
ideal.  Of our 134 candidates, only 11 stars had ${\rm [Fe/H]}<-1.5$
and MgH lines.  

For this search, we again utilized the reduced proper-motion
criterion to identify metal-poor stars drawn from the
NLTT catalog.  Instead of relying upon
Luyten's original photometry, we made use of the excellent work by
\citet{salim02,salim03} and \citet{gould03}.  Salim \& Gould cross-referenced the NLTT 
catalog with 2MASS \citep{2mass} and USNO-A \citep{usno} to
obtain optical and infrared photometry.  Proper-motions were
determined from the difference in positions between the two
catalogs.  \citet{salim02} then constructed an optical-infrared
reduced proper-motion diagram from which they identify ``distinct
tracks for white dwarfs, subdwarfs, and main-sequence stars''.  A
reduced proper-motion diagram based on the original Luyten photometry
does not show the expected separation between subdwarfs and main-sequence
stars.  Salim \& Gould suggest that the reduced proper-motion diagram
based on the original Luyten photometry suffers from a ``short color
baseline and large errors''.  
In Paper I, we confirmed the \citet{ryan89} statement that the photometric
``values tabulated in the NLTT catalog must be regarded as 
approximate only''.  Another advantage of the optical-infrared
reduced proper-motion diagram is that the colors provide an accurate 
estimate of the effective temperature.  
In Figure \ref{fig:empty}, we show a reduced proper-motion diagram
identical to that of Salim \& Gould with one exception.  Rather
than using $V+5\log\mu$, we use $J+5\log\mu$ as 
the reduced proper-motion.  The rationale is
that the 2MASS photometry is more accurate than USNO.  
Using photographic plates, the principal objective of the USNO catalog
is astrometry while the photometry is a by-product.  
USNO-A provides $B_U$ and $R_U$ magnitudes which are converted to $V$ 
according to the relation given in \citet{salim00} $V=R_U+0.23
+0.32(B-R)_U$.  \citet{salim03} show that the uncertainties in $V$ are
around 0.25 mag while uncertainties in $J$ are around 0.02 mag.  
However, there are no discernable differences between the original ($V+5\log\mu$)
reduced proper-motion diagram and ours ($J+5\log\mu$).
The locations of 
white dwarfs, subdwarfs, and main-sequence stars (solar metallicity) 
can be easily estimated.  

Of the stars we observed in Paper I, a subset had optical and 
infrared photometry from \citet{salim02,salim03}.  After constructing a reduced proper-motion
diagram, we overplotted the stars we observed in Paper I.  It was
clear that the metal-poor stars were separated from the metal-rich stars.
For this search, we selected stars with $H_J\ge4+2\times(V-J)$ where
$H_J=J+5\log\mu$ with $V$ and $J$ taken from \citet{salim02,salim03}.
This cutoff was the approximate line dividing
the subdwarfs from the main sequence stars studied in Paper I.
To ensure we observed cool stars, we generally
observed targets which had $1.4 < V-J < 2.6$.  The magnitude
limit was $V \le 13.9$, the limit of the McDonald 2.7m telescope.
We also imposed a declination limit $-40^o \le \delta \le 90^o$
appropriate for McDonald observatory.  This resulted in a list
of 470 candidate subdwarfs.  

In Paper I, the most metal-poor star we observed which showed MgH
lines was G39-36, \teff=4200K and [Fe/H]=$-2.5$.  Not 
only did this star have a large reduced
proper-motion, it was also marked by a large ultraviolet excess, 
$\delta(U-B)_{0.6}=0.25$ \citep{sandage86}.  \citet{roman55}
found a correlation between the ultraviolet excess and the strength
of metallic lines in high velocity stars.  \citet{ssh55} showed
that a lower metal abundance and the corresponding reduction in line
blanketing was responsible for the ultraviolet excess (see 
also \citealt{carney79} and references within).  Therefore, we also observed
cool stars with large ultraviolet excesses taken from 
\citet{sandage86}.  Finally, bright solar metallicity 
stars expected to have strong MgH lines
were also added to the list of candidates to be observed.  
These bright stars
were taken from the \hipparcos catalog \citep{hipparcos} and
from \citet{carney94}.  In total we observed 91 targets; 58 selected from the
$V-J$ reduced proper-motion diagram, 24 bright targets, and 9 
from \citep{sandage86}.  

\section{Observations and data reduction}
\label{sec:data}

Table \ref{tab:param} contains the list of candidates observed
at McDonald Observatory on the 2.7m Harlan J. Smith telescope 
between September 2002 and February 2003.  The data were obtained
using the cross-dispersed echelle spectrometer \citep{tull95}
at the coud\'{e}~f/32.5 focus with a resolving power of either
30,000 or 60,000.  The detector was a 
Tektronix CCD with 24 $\mu{\rm m}^2$ pixels
in a $2048 \times 2048$ format.  Using this setting, the 
spectral coverage was from 3800\AA~to 
8900\AA~with gaps between the orders beyond 5800\AA.  These
observations incorporated the MgH $A-X$ lines near 5140\AA. 
Visual inspection of the data indicated the presence or
absence of MgH molecular features.  Of our sample of 91 stars,
14 stars did not show MgH lines and 77 stars 
showed MgH lines.  Numerous Fe {\scshape i}
and Fe {\scshape ii} lines were available in the observed
spectra for spectroscopic determination of the stellar parameters.
For each star, exposure times were generally 20-30 minutes
and multiple exposures were taken and co-added to increase the
signal-to-noise ratio when necessary.  Although varying 
from star to star, the typical signal-to-noise ratio
of the extracted one dimensional spectra was 60 per pixel at 6500\AA.
One dimensional wavelength calibrated normalized spectra
were extracted in the standard way using the
IRAF\footnote{IRAF is distributed by the National Optical Astronomy Observatories,
which are operated by the Association of Universities for Research
in Astronomy, Inc., under cooperative agreement with the National
Science Foundation.} package of programs.  Equivalent widths were measured
using IRAF where in general Gaussian profiles were fit to the observed profile.

\section{Analysis}
\label{sec:analysis}

\subsection{Deriving stellar parameters}

\begin{figure}
\epsscale{0.9}
\plotone{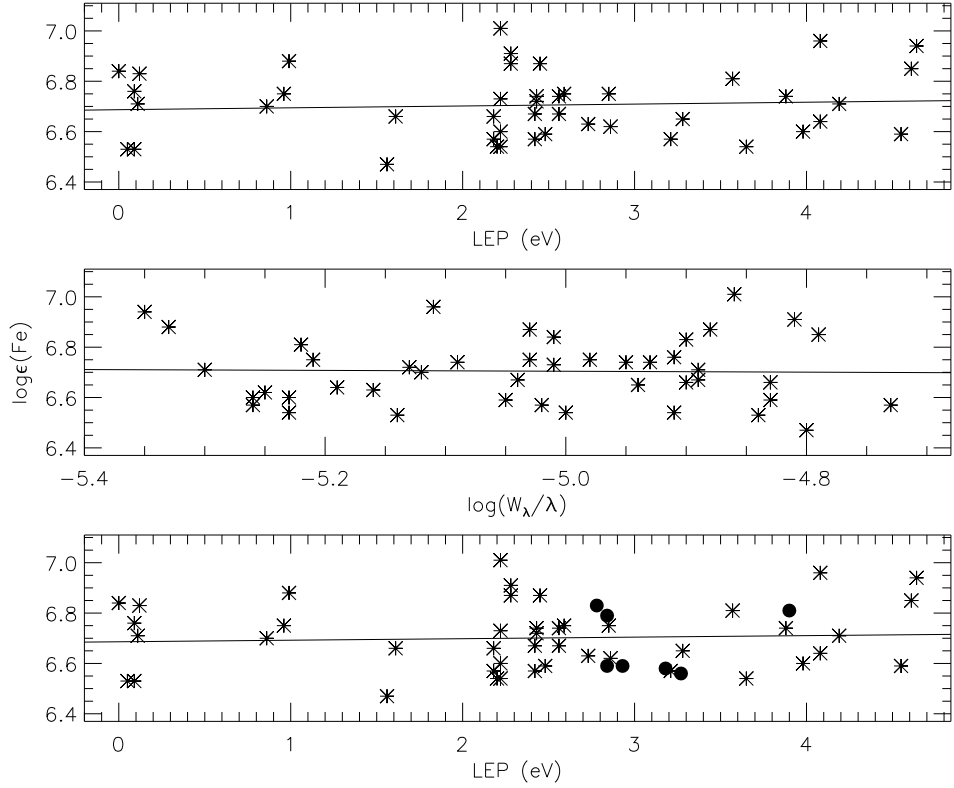}
\caption{Determination of stellar parameters \teff, log g, and $\xi_t$ using
excitation and ionization equilibria for G266-8.  In the top panel, the lower excitation
potential (LEP)-abundance relation is used to set \teff.  In the middle panel,
the reduced equivalent width (W$_\lambda/\lambda$)-abundance relation is used
to determine $\xi_t$.  In the bottom panel, the abundances of Fe {\scshape i} (asterisks)
and Fe {\scshape ii} (filled circles) are used to fix log g.  In all panels the line
represents the linear least squares fit to the data. \label{fig:param}}
\end{figure}

We determined the stellar parameters using the same method described in
Paper I.  The local thermodynamic equilibrium (LTE) stellar line analysis program 
{\scshape Moog} \citep{moog} was used in
combination with the adopted model atmosphere.
For log g $>$ 3.5, we used 
the {\scshape Nextgen} model atmosphere grid for
low mass stars computed by \citet{nextgen99}.  For giants with
log g $\le$ 3.5 we used LTE model atmospheres computed by \citet{kurucz93}.
In both cases we interpolated within the grid when necessary.
Equivalent widths (EWs) of 35 Fe {\scshape i} lines and 5 Fe {\scshape ii} lines
were measured.  The $gf$ values of the lines were taken from \citet{lambert96} and
from a compilation by R.E. Luck (1993, private communication).
The effective temperature (\teff) was set by insisting that the abundances of 
individual Fe lines be independent of lower excitation potential.  
Ideally, the \citet{alonso96b,alonso99b} {\em \teff:[Fe/H]:color} relations
based on the infrared flux method would be used to obtain values for \teff.
In particular, the $(V-K)$ color provides the greatest accuracy.  Unfortunately,
for [Fe/H]$\le-1.5$, the \citet{alonso96b} $(V-K)$ relation is only applicable above 
\teff$\simeq$4600K and \citet{salim03} have shown that errors in $V$ are 0.25 mag.
The microturbulence ($\xi_t$) was set by requiring that the abundances 
of individual Fe lines show no trend against equivalent width.  By forcing agreement
between the Fe abundance derived from neutral and ionized lines, the gravity (log g) was
fixed.  This process required iteration until a consistent set of parameters
were obtained (see Figure \ref{fig:param}) from which 
the Fe abundance was determined from the mean of
all Fe lines.  

Only weak lines, EWs $<$ 90 m\AA, were used in the analysis if possible.  We observed
the solar spectrum at R=60,000 to check our analysis techniques.  We measured
30 Fe {\scshape i} lines and 7 Fe {\scshape ii} lines and 
compared our equivalent widths with the \citet{grevesse99} values.
Our equivalent widths were larger by a mean value of 3.7 m\AA~with a standard deviation
of 2.6 m\AA.  We note that our equivalent widths were measured from a disk integrated
solar spectrum whereas the \citet{grevesse99} equivalent widths were measured
at disk center from the \citet{delbouille73} solar atlas.  
Using our equivalent widths and a {\scshape Nextgen} model atmosphere, we derived
a solar abundance of log$\epsilon$(Fe)$=7.54$.  Considering the \citet{grevesse99}
value of log$\epsilon$(Fe)$=7.50\pm 0.05$ derived from their empirical model solar
atmosphere, we adopted log$\epsilon$(Fe)$=7.52$
as the solar Fe abundance for this study.

The derived model parameters have uncertainties which were
estimated in the following way.  We varied \teff~until the
trends between lower excitation potential and abundance
were unacceptable.  Likewise, we changed the microturbulence
until the trends between equivalent width and abundance
were poor.  The gravity was adjusted until the difference between
the abundance from Fe {\scshape i} and Fe {\scshape ii} lines was
greater than the standard deviation of the Fe abundance derived
from Fe {\scshape i} lines (typically 0.1$-$0.15 dex).
Estimated uncertainties in the model parameters are
$\delta$\teff=150K, $\delta$log g=0.3 dex, 
$\delta\xi_t$=0.3 km s$^{-1}$, and $\delta$[Fe/H]=0.2 dex.  Though
the targets were fainter than those in Paper I, we increased the
number of exposures to ensure the signal-to-noise ratio was similar.
And so the uncertainties in the model parameters are identical
to those in Paper I.

\subsection{Comparison with literature}

\begin{figure}
\epsscale{0.8}
\plotone{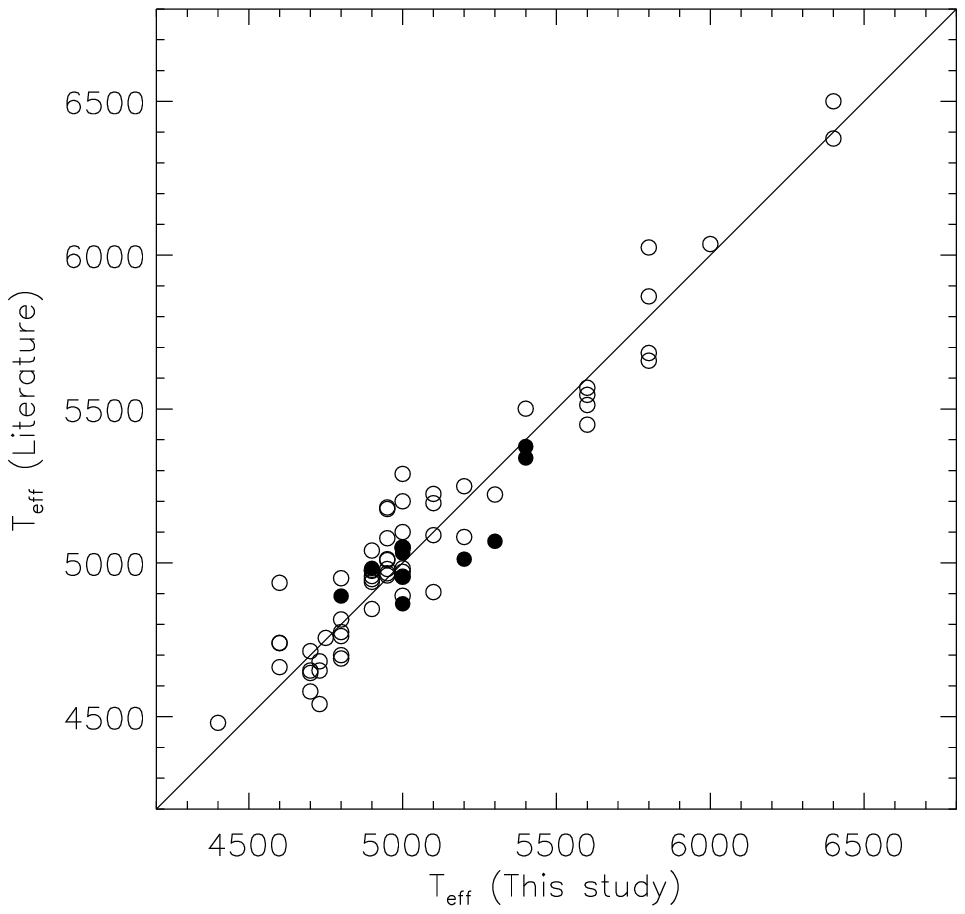}
\caption{\teff~comparisons between Yong \& Lambert (this study
and Paper I) and various literature studies.  The open circles 
represent the data from Paper I while the filled circles are stars
from this study.  The solid line represents the line of equality.
\label{fig:comp.teff}}
\end{figure}

\begin{figure}
\epsscale{0.8}
\plotone{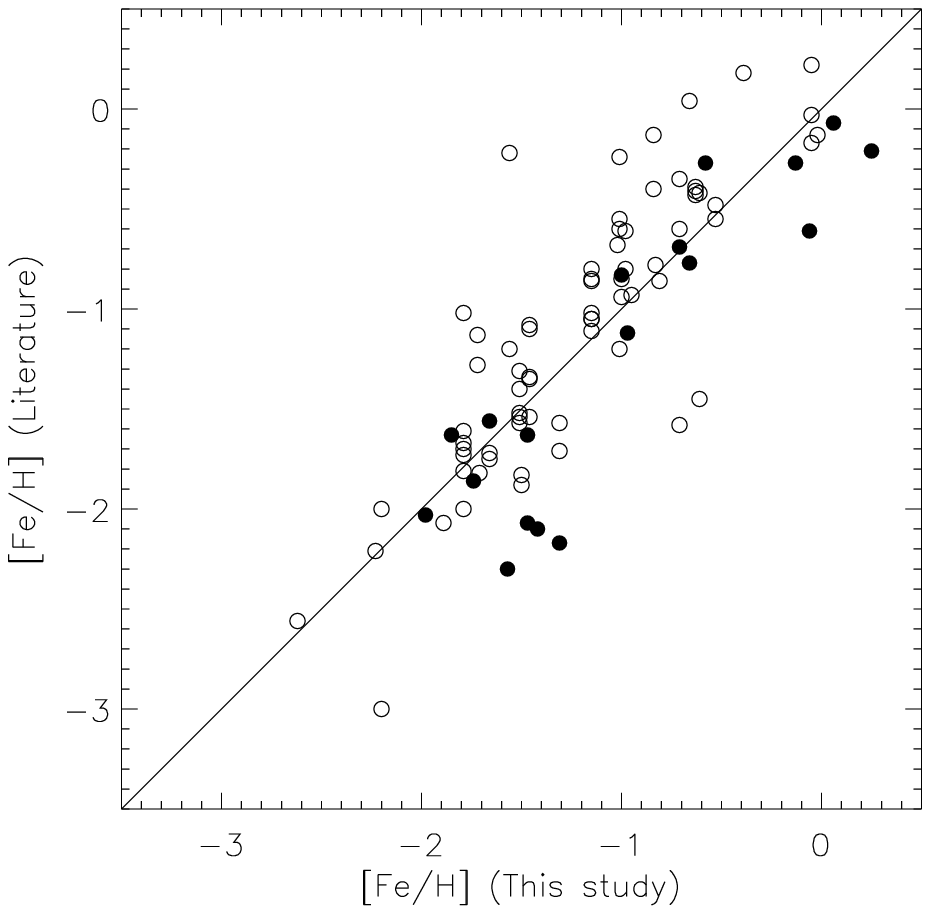}
\caption{[Fe/H] comparisons between Yong \& Lambert (this study
and Paper I) and various literature studies.  The open circles
represent the data from Paper I while the filled circles are stars
from this study.  The solid line represents the line of equality.
\label{fig:comp.fe}}
\end{figure}

A search on SIMBAD indicated that 10 stars had previously determined temperatures
and 17 stars had previously determined metallicities.  In Table \ref{tab:comp}, we compare
our values with those found in the literature.  Combining our data with Paper I,
we find a mean offset $\langle$\teff (Yong \& Lambert)$-$\teff (literature)$\rangle$ 
$= -16$ K with a standard deviation of 115K (see Figure 
\ref{fig:comp.teff}).  Again combining our data with Paper I,
the mean offset is $\langle$[Fe/H] (Yong \& Lambert)$-$[Fe/H] (literature)$\rangle = -0.06$ dex
with a standard deviation of 0.38 dex (see Figure \ref{fig:comp.fe}).
There is a reasonable agreement 
between the stellar parameters derived in this study (and Paper I) and the
values found in a variety of sources in the literature.

\subsection{Self-consistency check}

\begin{figure}
\epsscale{1.0}
\plotone{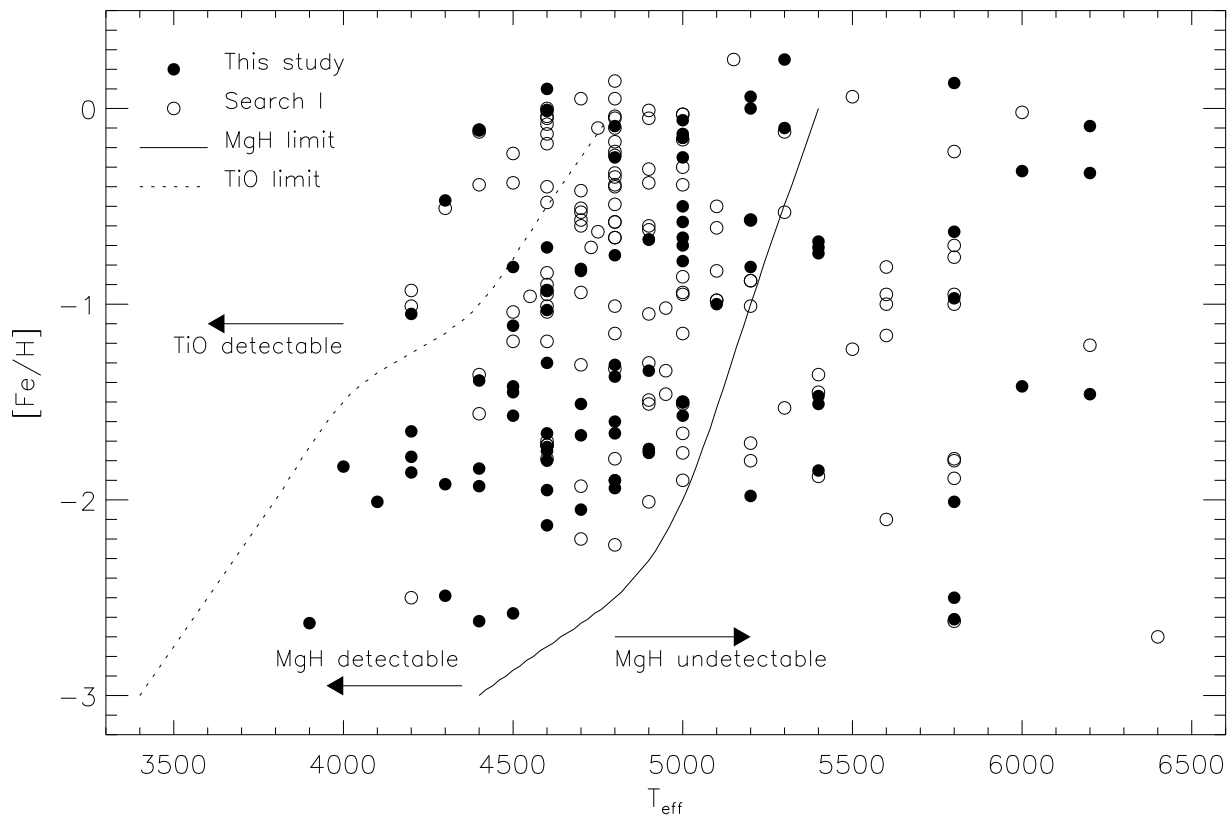}
\caption{Limit of detection for MgH and TiO features based on spectral synthesis
of representative lines assuming log g = 4.5 and R=60,000.  The filled circles
are from this study while the open circles are from Paper I.  With the exception of
a few stars from Paper I, all stars with MgH lines lie to the left of the 
MgH detection limit.  Also, all stars without MgH lines lie to the right of the MgH
detection limit.  The few stars that do not lie where they ought to are giants
or were observed at lower resolving powers.
\label{fig:detect}}
\end{figure}

Molecular lines will not be visible in the spectra of stars
with sufficiently high temperatures and/or low abundances.  
A simple self-consistency test of our stellar parameters is to predict the limit
in temperature and metallicity at which molecular lines will not be observed.  Once
a detection limit is obtained, we can then check the stellar parameters to
determine whether molecular lines should or should not be present.
We synthesized representative MgH and TiO lines
assuming a resolving power of 60,000 and log g $=4.5$.  The MgH molecular
data were taken from \citet{gl2000} while the TiO molecular data
were taken from \citet{jorgensen94}.  Values 
of [Ti/Fe] and [Mg/Fe] in accord with observed trends summarized in
\citet{alc01} were assumed, i.e., [Mg/Fe]=0.4 at [Fe/H]=$-1.5$.  At a given
value for \teff, we decreased the metallicity until the molecular
features reached a depth of $\sim$5\% relative to the continuum.  We considered this
value of \teff~and [Fe/H] to be our detection limit.  
In Figure \ref{fig:detect}, we plot our stars
in temperature--metallicity space imposing our limits of detection for
MgH and TiO.  At a fixed metallicity, MgH is detectable to considerably
higher values of \teff~than TiO.  Despite TiO (6.9 eV) having a higher
dissociation energy than MgH (1.34 eV), the overwhelming H abundance 
ensures the continuing presence of MgH to higher temperatures.

All 14 stars that show neither MgH nor TiO lines occupy a region in 
temperature-metallicity space where, for the given stellar parameters, 
we would not expect to see molecular lines.
All 77 stars that show MgH lines lie in the region where, for the given
stellar parameters, we do expect
to see molecular lines.  Our approximate limits of detection assume
a resolving power of 60,000 and log g = 4.5.  For objects
observed at a lower resolving power, we expect the detection limit 
would move to lower values of \teff~at a fixed metallicity.  Likewise,
the detection limit for giants would be shifted to lower values of \teff~at a 
fixed metallicity.
The vast majority of the stars observed in Paper I also lie in the expected regions.  That is,
stars with MgH lines lie to the left of the detection limit while stars without
MgH lines fall to the right of the detection limit in Figure \ref{fig:detect}.
We note that the handful of stars that do not conform
to the predictions are observed at lower resolving powers or are giants.
Through the synthesis of representative molecular lines, we 
have performed a self-consistency check validating our derived stellar parameters.

\section{Discussion}
\label{sec:discussion}

\subsection{Kinematics}

\begin{figure}
\epsscale{0.8}
\plotone{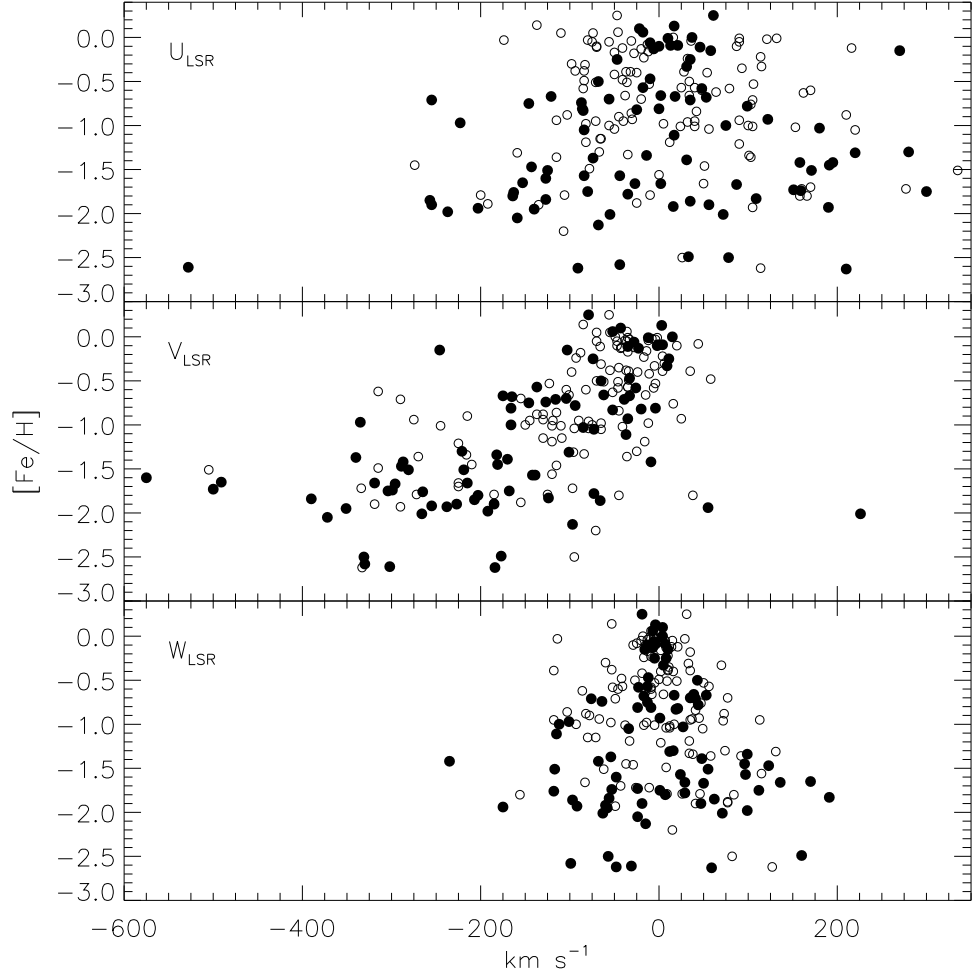}
\caption{Galactic space-velocity U, V, and W (relative to the LSR) vs [Fe/H].
The open circles represent the data from Paper I while the filled circles are stars
from this study.  Only stars which have
$\frac{\sigma_U+\sigma_V+\sigma_W}{|U|+|V|+|W|}<0.7$ are 
shown.  A considerable fraction
of the sample noticeably lag the LSR (V $<-50$ km/s).
\label{fig:uvw}}
\end{figure}

We used the reduced proper-motion constraint to select stars kinematically
distinct from the thin disk.  This selection criterion was used in order to observe stars
on halo-like orbits with halo-like metallicities.  The first check we can implement
is to determine whether we successfully targeted stars with kinematics
unlike thin disk stars.  Following the 
\citet{johnson87} recipe, we calculated the Galactic space-velocity 
components U (positive towards the Galactic center), 
V (positive in the direction of Galactic rotation), and 
W (positive towards the north Galactic pole)
along with the associated uncertainties $\sigma_U$, $\sigma_V$, and $\sigma_W$ 
(see Table \ref{tab:param}).
For the solar motion with respect to the local standard of rest (LSR),
we assumed the \citet{lsr} values (+10,+5,+7) km s$^{-1}$ in (U,V,W).
Note that in Paper I, we incorrectly applied a correction of $-$10 km s$^{-1}$
in U.  Therefore, +20 km s$^{-1}$ must be added to our values of U in Paper I
to obtain U$_{LSR}$.  In the absence of \hipparcos parallaxes, 
spectroscopic parallaxes were determined by
using the derived model parameters and the Y$^{2}$ isochrones \citep{yi01}.
In Figure \ref{fig:uvw} we plot U$_{LSR}$, V$_{LSR}$, and W$_{LSR}$ versus
[Fe/H] including the stars from Paper I.  As in Paper I, we identify
stars that lag the LSR, V $<-$50 km s${-1}$, as likely members of the
thick disk or halo.  As expected, our reduced proper-motion constraint enabled 
us to select stars belonging to populations kinematically distinct from the thin disk.  

\subsection{Metallicity}

Our goal was to find stars with [Fe/H]$\le -1.5$.  The second check we
can conduct is to verify the success of our selection criteria in targeting
metal-poor stars.
In Figure \ref{fig:hist.fe} we plot the number of stars versus metallicity
for this study, Paper I, \citet{carney94}, and the \citet{ryan91} sample.
In Paper I, we noted that our distribution was similar to the Carney sample.
We also mentioned that the Ryan sample had a peak at lower metallicity
since these stars were observations of metal-poor candidates identified
by \citet{ryan89} as having an ultraviolet excess 
corresponding to [Fe/H]$<-1.2$, that is, $\delta(U-B)_{0.6}>0.2$.  In this study,
we were far more successful in identifying metal-poor stars with
38 out of 91 stars being more metal-poor than 
[Fe/H]$=-1.5$.  As a comparison, in Paper I we had 27 of
134 stars with [Fe/H]$\le -1.5$.  In this study, 32 stars were more
metal-poor than [Fe/H]$=-1.5$ with MgH lines present in 
their spectra.  As a comparison, 11 stars in Paper I were more metal-poor
than [Fe/H]$=-1.5$ with MgH lines present in 
their spectra.  Clearly the selection technique utilized in this paper was
more successful than the technique applied in Paper I.  That is, the
\citet{salim02} optical-infrared reduced proper-motion diagram separates
subdwarfs from main-sequence stars more effectively than a reduced proper-motion diagram
based on the original Luyten photometry.

Let us consider only those stars with 2MASS and USNO photometry, i.e.,
only the stars in Figure \ref{fig:vmj}.  After combining
this study with the stars in Paper I, we find that 56\% (43 of 76)
of stars that lie below the cutoff have [Fe/H]$\le -1.5$.  While this is
far from the ultra metal-poor stars currently being discovered by other
programs, [Fe/H]$\le -1.5$ is a metallicity regime in which the 
Mg isotope ratios have not been explored.  In this metallicity range,
only cool subdwarfs will show MgH lines.  We also note that beyond
$J+5\log{\mu}=10$, all of the observed stars are more metal-poor
than [Fe/H]$=-1.5$.

\subsection{Temperature}

\begin{figure}
\epsscale{0.6}
\plotone{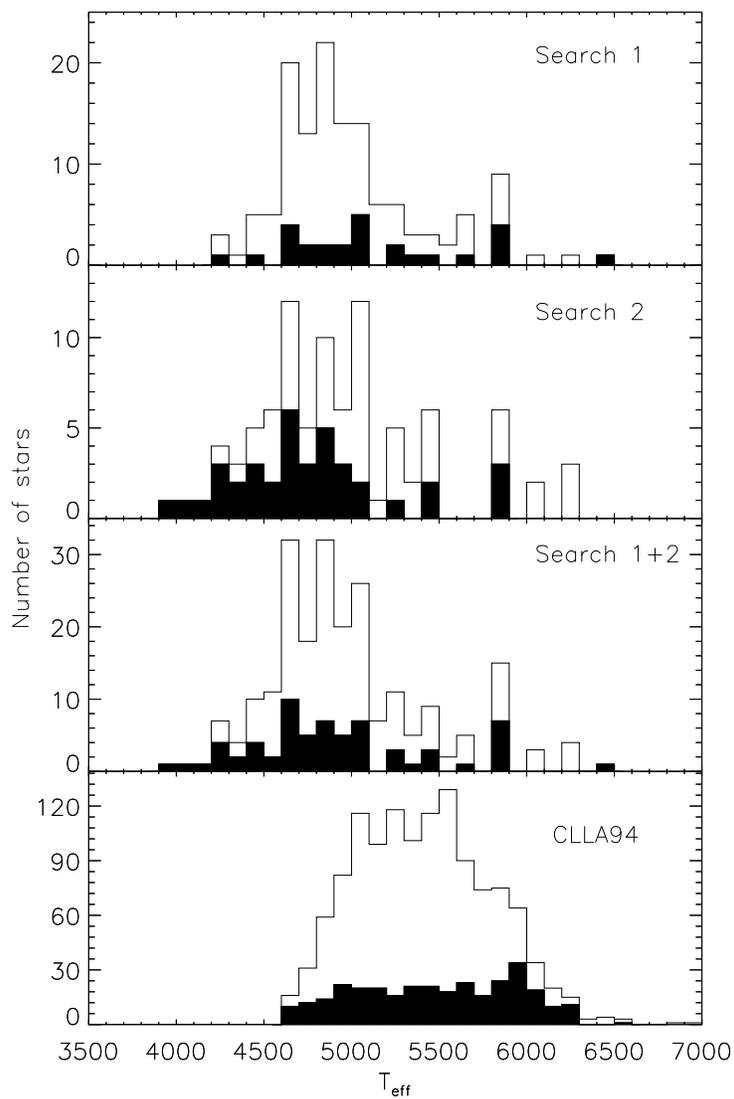}
\caption{Number of stars versus effective temperature for Paper I (top panel),
this study (2nd panel), Paper I and this study (3rd panel), and the \citet{carney94} 
study (bottom panel).  The filled histograms represent the distribution of stars
with [Fe/H]$\le-1.5$.  Clearly our samples peak at lower values of \teff~than the
Carney sample.  The 2nd panel shows that our selection technique is very efficient
at identifying cool stars with [Fe/H]$\le-1.5$.
\label{fig:hist.teff}}
\end{figure}

Our goal was to find stars with \teff$<$4700K.  More specifically, ideal targets 
were stars with [Fe/H]$<-1.5$ with MgH 
molecular lines.  Of the 91 stars presented in this paper, 38 had
\teff$\le$4700K and given the uncertainties in our derived effective
temperatures, we note that 48 of the 91 stars had \teff$\le$4800K.  As a comparison,
in Paper I 44 of the 134 stars had \teff$<$4700K and 69 of the 134 stars had
\teff$\le$4800K.  In Figure
\ref{fig:hist.teff}, we plot the number of stars versus \teff~for this study,
Paper I, and the \citet{carney94} sample.  Our temperature distributions
are similar between this study and Paper I.  Both of our searches have
focused upon stars considerably cooler than the Carney study which highlights
the different temperature regimes of interest.  Importantly, 77 of the
91 stars presented here have MgH lines.  In Paper I, 100 of the 134 stars
had MgH lines.

\begin{figure}
\epsscale{0.8}
\plotone{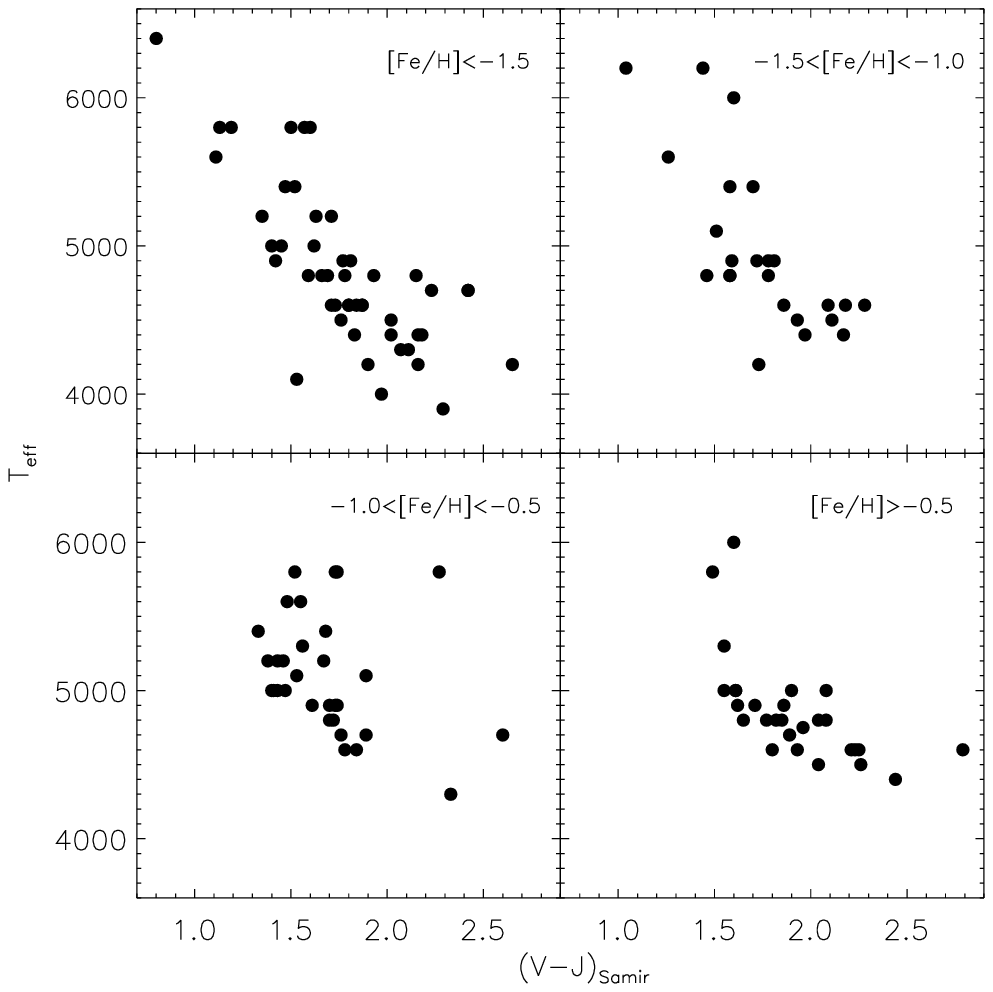}
\caption{$V-J$ color from \citet{salim02} versus effective temperature
derived from this study and Paper I.  In the upper left panel are stars 
with [Fe/H]$< -1.5$, in the upper right panel are stars with $-1.5<$[Fe/H]$<-1.0$, 
in the lower left panel are stars with $-1.0<$[Fe/H]$<-0.5$,
and in the lower right panel are stars with [Fe/H]$> -0.5$.  Given a value of
$V-J$, the corresponding \teff~can be determined with reasonable accuracy.
\label{fig:teff.vj}}
\end{figure}

\begin{figure}
\epsscale{0.8}
\plotone{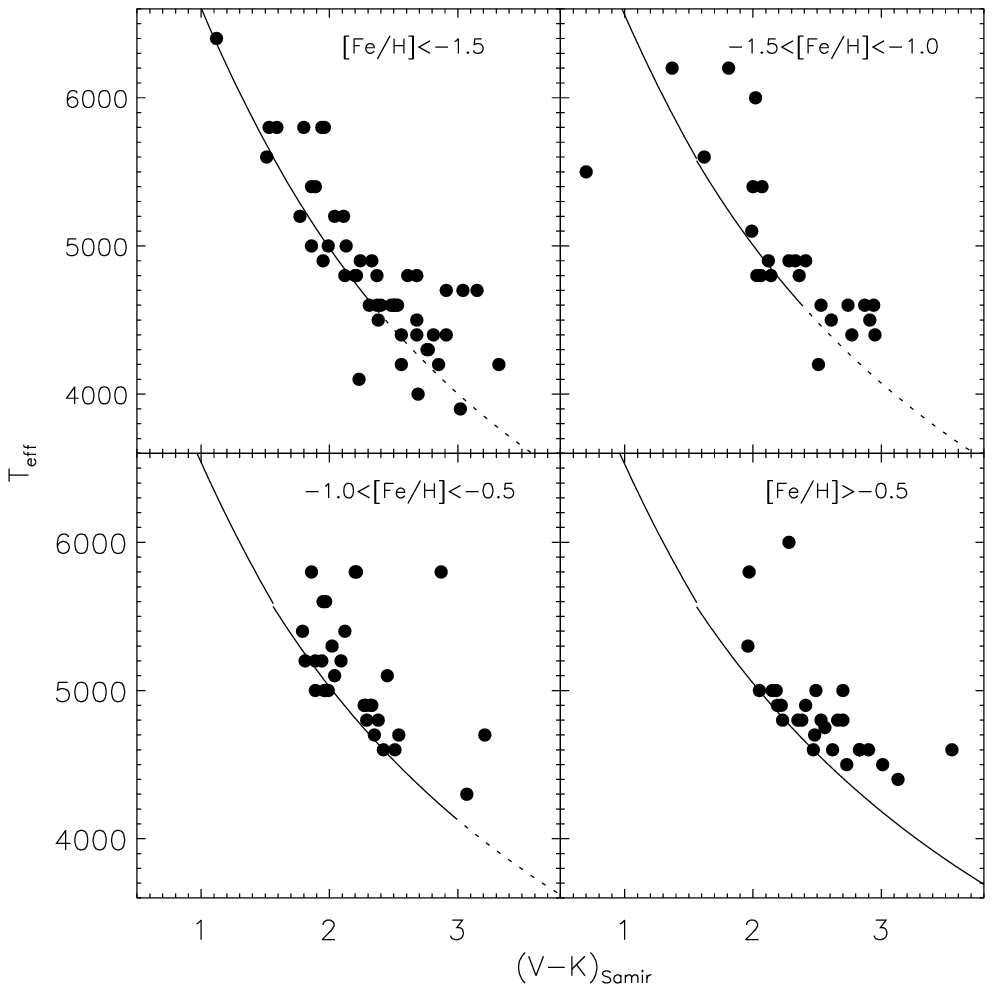}
\caption{$V-K$ color from \citet{salim02} versus effective temperature
derived from this study and Paper I.  In the upper left panel are stars 
with [Fe/H]$< -1.5$, in the upper right panel are stars with $-1.5<$[Fe/H]$<-1.0$, 
in the lower left panel are stars with $-1.0<$[Fe/H]$<-0.5$,
and in the lower right panel are stars with [Fe/H]$> -0.5$.  In each panel, the
solid line is the $(V-K)$:\teff~relation from \citet{alonso96b} for dwarfs.  The
dotted line is the extrapolation of the $(V-K)$:\teff~relation.
\label{fig:teff.vk}}
\end{figure}

In Paper I, we confirmed Ryan's findings that the Luyten colors were inaccurate.
Firstly, the magnitudes were published to only 0.1 mag.  Secondly, we chose stars
assigned by Luyten to color classes g-k, k, or k-m.  These stars spanned 
4200K $<$ \teff $<$ 6400K yet the Luyten
colors for this range of stars were $m_{\rm pg}-m_R$=0.9, 1.1, or 1.3.
We now comment on the colors in the \citet{salim02} study.  In Figure
\ref{fig:teff.vj}, we plot the $V-J$ color against our derived temperatures (this study
and Paper I) separated into four metallicity bins.  This Figure shows
that a fair estimate of the effective temperature can be obtained from the
$V-J$ color.
\citet{alonso96b,alonso99b} derived {\em \teff:[Fe/H]:color}
relations for dwarfs and giants using the infrared flux method.  
They have shown that the
$V-K$ color index provides the most accurate temperature estimates.  
(We refrain from using the infrared flux method calibrations due to the uncertainties in
$V$.) In Figure
\ref{fig:teff.vk}, we plot the $V-K$ color against our derived temperatures
and find that the dispersion about the mean is slightly less than for
$V-J$.  The $K$ magnitudes were also from 2MASS and presented by \citet{salim02}.
This reduced scatter in the $V-K$ versus \teff~plot suggests
that a reduced proper-motion diagram featuring 
$V-K$ versus $K+5\log{\mu}$ may be a power tool when the metal-poor
targets need to span a small temperature range.  In this Figure we also
plot the \citet{alonso96b} $(V-K)$:\teff~relation for dwarfs.  We find that
this relation reasonably matches our data though at the metal-poor end,
our data extend to cooler temperatures than the \citet{alonso96b} relations.
In both the $V-J$ and $V-K$ plots (Figures \ref{fig:teff.vj} and \ref{fig:teff.vk}),
we suspect that the dispersion is primarily due to errors in V.  Earlier
we mentioned that the errors in $V$ were around 0.25 mag while the 
errors in $J$ (and presumably $K$) were only 0.02 mag \citep{salim03}.
While reddening may also be a factor, we note that these cool dwarfs 
lie within 100pc of the sun and errors in $V-J$ and $V-K$ would overwhelm
any reddening effects.

\begin{figure}
\epsscale{1.0}
\plottwo{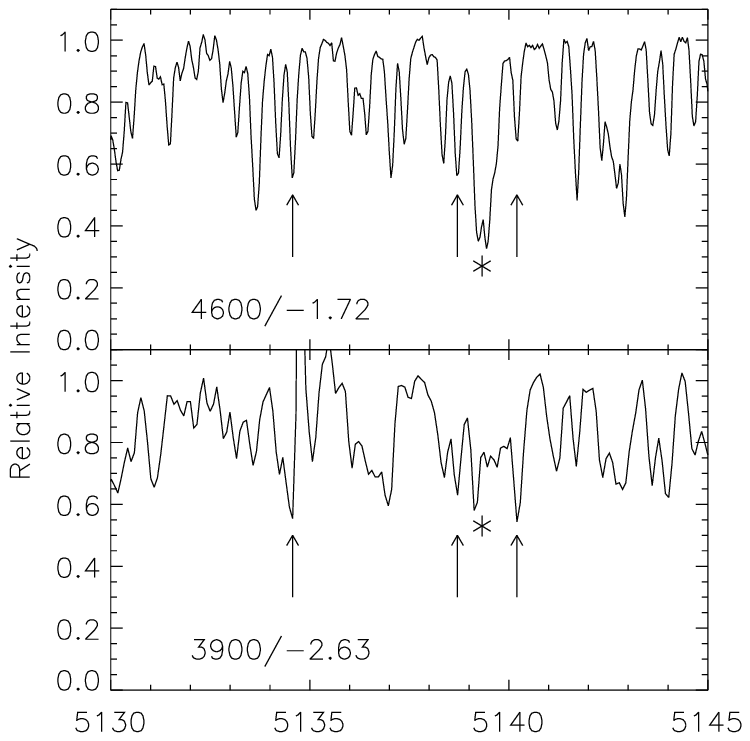}{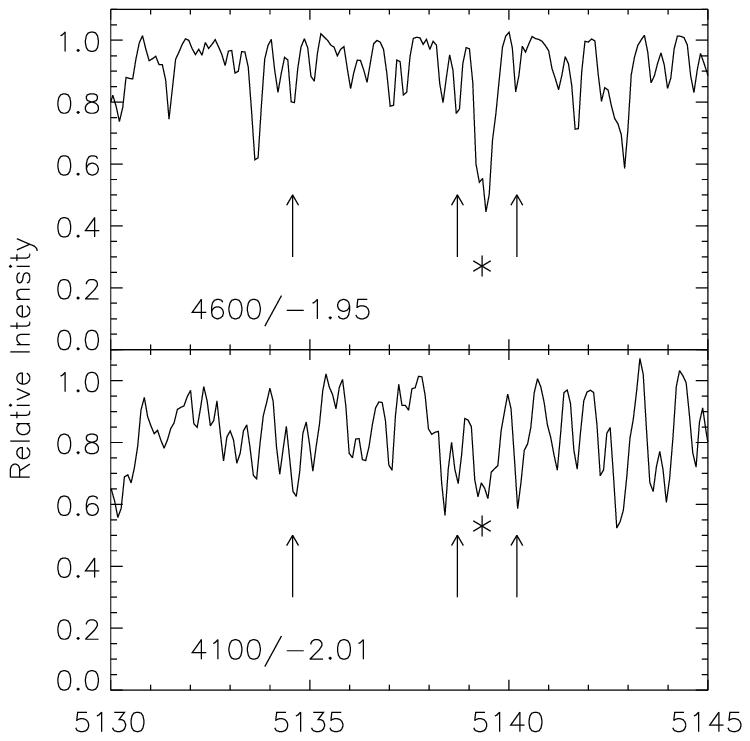}
\caption{Observed spectra of 4 different cool subdwarfs.  The effective temperature
and metallicity of each star is shown.  In each panel, the 3 arrows highlight
representative MgH lines while the asterisk highlights an atomic line (Fe and Ni).
In the upper panels, the 3 MgH lines are dwarfed by the strong atomic line.  In
the lower panels, the 3 MgH lines are of comparable strength to the atomic line.
\label{fig:mgh}}
\end{figure}

Based on spectrum synthesis, \citet{cottrell78} showed that MgH lines in
cool stars can be a useful metal abundance discriminant.  In cool stars, the strength
of the MgH lines does not strongly depend on metallicity, unlike atomic lines.  For
sufficiently cool stars, a decrease in the metal abundance will weaken the atomic
lines whilst the MgH lines remain strong.  In Figure \ref{fig:mgh}, we show
the spectra of 4 stars highlighting some MgH lines and a strong atomic line.
In 2 of the spectra the ``normal'' situation is shown in which the MgH lines are 
dwarfed by the strong atomic line.  In the other 2 spectra, Cottrell's prediction
is verified since the MgH lines
are comparable in strength to the atomic line.  That is, the atomic line has weakened
while the MgH lines have maintained their strength.  Note that all stars shown
in Figure \ref{fig:mgh} are very cool.  If these stars were solar metallicity, 
Figure \ref{fig:detect} shows that the spectra of these stars would be swamped 
with TiO lines.

\section{Concluding remarks}
\label{sec:remarks}

We present stellar parameters for 91 candidate subdwarfs selected primarily
by their large reduced proper-motions.  Our goal was to identify metal-poor
stars [Fe/H]$<-1.5$ with MgH lines.  This is the 2nd search
that we have conducted for these cool subdwarfs.  In 
our first search, we drew candidates from a reduced proper-motion 
diagram based on Luyten's original photometry.  While we were able to
find cool subdwarfs, our efficiency was low (11 of 134 stars had [Fe/H]$<-1.5$ 
and MgH lines).  Errors in the
Luyten photometry resulted in a reduced proper-motion diagram without
cleanly separated populations of subdwarfs and solar-metallicity main-sequence stars.
In this search, we
made use of the 2MASS infrared and USNO optical photometry presented
by \citet{salim02}.  The reduced proper-motion diagram based on this
photometry showed distinct populations of subdwarfs and main-sequence stars.
Spectroscopic observations of these candidate subdwarfs showed that the majority
were metal-poor stars.  
Of the candidate subdwarfs selected from their location in the 
optical-infrared reduced proper-motion diagram,
we have shown that 56\% have [Fe/H]$<-1.5$.
In Paper I, only 20\% of candidates had [Fe/H]$<-1.5$.
Further, we verified that the $V-J$ color index was well correlated with \teff~such that
cool stars suspected of having MgH lines could be targeted.  We intend
to make further observations to identify cool subdwarfs.  In this search, we found
a further 32 stars with [Fe/H]$<-1.5$ and MgH lines.  We 
intend to re-observe candidates from this study with higher resolving
power and higher signal-to-noise ratio in order to measure the Mg
isotope ratios to study Galactic chemical evolution.

\acknowledgments

We are grateful to Samir Salim and Andy Gould for providing a list of
candidate subdwarfs prior to publication.  
We thank G. Fritz Benedict for valuable discussions and
the referee, Andy Gould, for helpful comments.
We acknowledge support from the Robert A. Welch
Foundation of Houston, Texas.
This research has made use of the SIMBAD database,
operated at CDS, Strasbourg, France and
NASA's Astrophysics Data System.

\clearpage

\begin{figure}
\epsscale{1.0}
\caption{$V-J$ reduced proper-motion diagram.  The points are NLTT stars
with 2MASS and USNO-A identifications make by \citet{salim02,salim03}.  
Main-sequence stars (MS) are separated from the subdwarfs (SD) by 
the solid line.  White dwarfs (WD) are separated from SDs and MS
stars by the dashed line.  
\label{fig:empty}}
\end{figure}

\clearpage

\clearpage

\clearpage

\clearpage

\clearpage

\clearpage

\begin{figure}
\epsscale{0.6}
\plotone{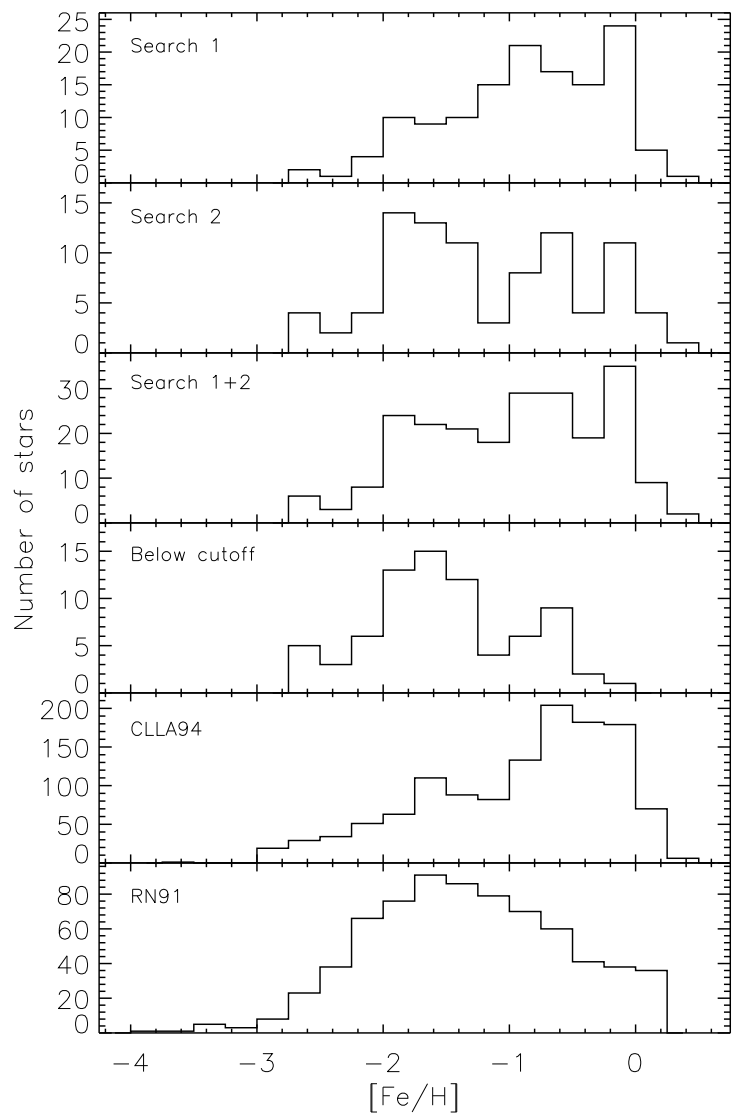}
\caption{Number of stars versus metallicity for Paper I (top panel),
this study (2nd panel), Paper I and this study (3rd panel), only
the stars falling below the cutoff defined in Figure \ref{fig:empty} (4th panel),
the \citet{carney94} study (5th panel), and the \citet{ryan91} study (bottom panel).
The 4th panel shows most clearly that our selection technique efficiently 
selects metal-poor stars.
\label{fig:hist.fe}}
\end{figure}

\clearpage

\begin{figure}
\epsscale{1.0}
\caption{$V-J$ reduced proper motion diagram based on the \citet{salim02} photometry.
In each panel, the small dots are stars in the NLTT catalog with 2MASS and USNO photometry.  
We overplot (large circles) the locations of stars we previously observed 
for which we were able to derive metallicities either in this study or in Paper I.
In the upper left panel the large circles are stars with [Fe/H]$< -1.5$, in the upper right panel
are stars with $-1.5<$[Fe/H]$<-1.0$, in the lower left panel are stars with $-1.0<$[Fe/H]$<-0.5$,
and in the lower right panel are stars with [Fe/H]$> -0.5$.
The dotted and dashed lines (based on Samir \& Gould) define approximate 
boundaries between main sequence stars (MS), subdwarfs (SD), and white 
dwarfs (WD).  In the upper left panel, the stars with [Fe/H]$< -2.0$ (open circles) 
overlap with the [Fe/H]$< -1.5$ stars.
\label{fig:vmj}}
\end{figure}

\clearpage

\clearpage

\clearpage

\clearpage

\clearpage

\input{tab1}

\input{tab2}

\end{document}

%% file: tab1.tex
\begin{deluxetable}{lcrccccrrrrrrcc} 
\tabletypesize{\tiny}
\tablecolumns{15} 
\tablewidth{0pc} 
\tablecaption{Basic data and derived parameters for objects \label{tab:param}}
\tablehead{ 
\colhead{Name}          & \colhead{Ra\tablenotemark{a}}     &
\colhead{Dec\tablenotemark{b}}           & 
\colhead{\teff} &
\colhead{[Fe/H]}        & \colhead{log g} &
\colhead{$\xi_t$}       & \colhead{U$_{\rm LSR}$} &
\colhead{$\sigma_U$}    & \colhead{V$_{\rm LSR}$} &
\colhead{$\sigma_V$}    & \colhead{W$_{\rm LSR}$} &
\colhead{$\sigma_W$}    & \colhead{Molecular} &
\colhead{Source} \\ 
\cline{7-13} \\
\colhead{}              & \colhead{2000}     &
\colhead{2000}          & 
\colhead{(K)} &
\colhead{}              & \colhead{(cm s$^{-2}$)} &
\multicolumn{7}{c}{(km/s)}  & \colhead{Features} &
\colhead{}
}
\startdata 
LP 644-47  & 820 & -51448 & 5400 & -1.47 & 4.5 & 0.8 & -163 & 83 & -294 & 93 & 116 & 38 &  N  &  Samir02 \\
NLTT 718  & 1445 & 412237 & 4800 & -0.75 & 4.5 & 0.2 & -166 & 109 & -151 & 61 & -20 & 21 &  Y  &  Samir02 \\
G 31-56  & 2947 & 10112 & 4600 & -0.93 & 4.5 & 0.3 & 102 & 36 & -40 & 20 & -6 & 15 &  Y  &  Sandage86 \\
LP 705-100  & 3613 & -84036 & 5400 & -0.74 & 4.5 & 0.8 & -107 & 63 & -132 & 81 & -71 & 18 &  Y  &  Samir02 \\
LP 705-89  & 4243 & -123918 & 5800 & -0.97 & 4.5 & 0.6 & -243 & 110 & -340 & 155 & -108 & 25 &  N  &  Samir02 \\
G 69-18  & 4432 & 295424 & 4500 & -2.58 & 4.5 & 0.3 & -64 & 44 & -335 & 70 & -106 & 58 &  Y  &  Samir02 \\
G 1-19  & 4449 & -642 & 4200 & -1.78 & 4.5 & 0.3 & -55 & 22 & -78 & 19 & 22 & 3 &  Y  &  Sandage86 \\
LP 882-323  & 10956 & -300853 & 4100 & -2.01 & 4.5 & 0.4 & -75 & 15 & 221 & 37 & 64 & 3 &  Y  &  Samir02 \\
HIP 6001  & 11711 & 93004 & 4600 & -0.71 & 4.5 & 0.3 & 15 & 4 & -44 & 5 & 34 & 1 &  Y  &  Hipp \\
G 272-38  & 13303 & -161100 & 4400 & -1.39 & 4.5 & 1.1 & 11 & 6 & -175 & 67 & 41 & 3 &  Y  &  Samir02 \\
G 72-18  & 13756 & 322100 & 4900 & -1.74 & 4.5 & 0.4 & 139 & 10 & -304 & 77 & -60 & 87 &  Y  &  Samir02 \\
G 3-13  & 14258 & 170802 & 4700 & -1.67 & 4.5 & 0.6 & 67 & 20 & -301 & 78 & 43 & 36 &  Y  &  Sandage86 \\
HIP 8775  & 15251 & 44958 & 4600 & -0.01 & 4.5 & 0.8 & -10 & 3 & -17 & 3 & -4 & 1 &  Y  &  Hipp \\
LP 829-11  & 20514 & -202942 & 4900 & -1.34 & 4.5 & 0.3 & -34 & 34 & -187 & 101 & 92 & 19 &  Y  &  Samir02 \\
LP 829-30  & 21348 & -253100 & 4600 & -1.73 & 4.5 & 0.8 & 131 & 50 & -505 & 170 & -31 & 17 &  Y  &  Samir02 \\
G 3-44  & 21350 & 155910 & 4200 & -1.65 & 4.5 & 0.4 & -173 & 90 & -496 & 134 & 163 & 16 &  Y  &  Samir02 \\
NLTT 7419  & 21426 & -245707 & 5000 & -1.57 & 4.5 & 0.5 & -64 & 17 & -144 & 36 & 17 & 11 &  Y  &  Samir02 \\
HIP 10682  & 21727 & 270814 & 5000 & -0.15 & 4.5 & 0.5 & 38 & 3 & -108 & 9 & 3 & 5 &  Y  &  Hipp \\
LP 885-44\tablenotemark{c}  & 21738 & -320430 & 4900 & -0.67 & 4.5 & 0.4 & -141 & 47 & -38 & 11 & 10 & 12.8 &  Y  &  Samir02 \\
G 74-21  & 22415 & 411348 & 4700 & -1.51 & 4.5 & 0.5 & -145 & 82 & -286 & 97 & 48 & 12 &  Y  &  Samir02 \\
LP 770-34  & 23130 & -153648 & 4500 & -1.45 & 4.5 & 0.3 & 171 & 35 & -186 & 65 & 89 & 10 &  Y  &  Samir02 \\
G 75-23  & 23321 & -64354 & 5400 & -1.51 & 4.5 & 0.6 & 151 & 73 & -224 & 106 & -124 & 48 &  N  &  Samir02 \\
G 75-32  & 23935 & -51436 & 5000 & -0.7 & 4.5 & 0.4 & -76 & 22 & -109 & 33 & 28 & 14 &  Y  &  Samir02 \\
LP 887-18  & 25827 & -290428 & 4400 & -1.93 & 4.5 & 0.6 & 170 & 63 & -243 & 65 & -99 & 3 &  Y  &  Samir02 \\
HIP 14803  & 31110 & 91147 & 4600 & 0.1 & 4.5 & 0.7 & -42 & 4 & -48 & 8 & -3 & 3 &  Y  &  Hipp \\
G 5-26  & 31753 & 233720 & 4800 & -1.6 & 4 & 0.5 & -147 & 59 & -580 & 214 & -55 & 22 &  Y  &  Samir02 \\
LP 772-56  & 32353 & -171822 & 4400 & -1.84 & 4.5 & 0.4 & -147 & 9 & -395 & 88 & -63 & 33 &  Y  &  Samir02 \\
HIP 16739  & 33519 & 54331 & 4500 & -0.81 & 4.5 & 0.6 & -20 & 1 & -9 & 3 & -31 & 2 &  Y  &  Hipp \\
LP 773-20  & 33903 & -192636 & 5000 & -0.78 & 4.5 & 0.6 & 79 & 18 & -99 & 38 & 37 & 5 &  Y  &  Samir02 \\
HIP 17451  & 34427 & -4114 & 5800 & 0.13 & 4.5 & 0.8 & -3 & 1 & -2 & 1 & -11 & 1 &  N  &  Hipp \\
G 39-1\tablenotemark{c}  & 40734 & 380430 & 5000 & -0.06 & 4.5 & 1.1 & -30 & 2 & -28 & 6 & -6 & 2 &  Y  &  CLLA94 \\
HIP 19814  & 41458 & -53749 & 5400 & -0.71 & 4.5 & 1 & -275 & 58 & -121 & 45 & -83 & 90 &  Y  &  Hipp \\
HIP 21380\tablenotemark{c}  & 43517 & 230243 & 5300 & -0.1 & 4.5 & 1.1 & -20 & 1 & -2 & 1 & -13 & 1 &  Y  &  Hipp \\
G 8-45  & 43749 & 194018 & 4200 & -1.05 & 4.5 & 0.7 & -104 & 1 & -78 & 15 & -41 & 3 &  Y  &  Samir02 \\
G 83-26  & 43816 & 125100 & 4500 & -1.57 & 4.5 & 0.6 & -104 & 9 & -147 & 32 & 90 & 28 &  Y  &  Sandage86 \\
G 83-46  & 45623 & 145510 & 4400 & -2.62 & 4.25 & 0.3 & -111 & 4 & -189 & 36 & -55 & 4 &  Y  &  Sandage86 \\
WT 1428  & 45729 & -235723 & 4900 & -1.76 & 4.5 & 0.4 & -183 & 4 & -270 & 24 & -125 & 22 &  Y  &  Samir02 \\
HIP 23113  & 45824 & 514935 & 6200 & -0.09 & 4.25 & 1 & -7 & 2 & -1 & 3 & 0 & 2 &  N  &  Hipp \\
HIP 23198  & 45932 & 484645 & 4300 & -0.47 & 4.5 & 0.8 & -30 & 2 & -38 & 4 & -19 & 2 &  Y  &  Hipp \\
HIP 24289  & 51245 & 41916 & 5000 & -0.66 & 4.5 & 1.1 & -18 & 1 & -67 & 10 & 32 & 8 &  Y  &  Hipp \\
HIP 24655  & 51726 & 260449 & 6200 & -0.33 & 3.5 & 2.1 & 11 & 2 & 5 & 3 & -3 & 3 &  N  &  Hipp \\
NLTT 14937  & 52211 & 351726 & 4900 & -1.75 & 4.5 & 0.3 & -100 & 11 & -168 & 78 & 1 & 7 &  Y  &  Samir \\
HIP 26864  & 54213 & 72407 & 4500 & -1.11 & 4.5 & 0.4 & -3 & 4 & -42 & 3 & -122 & 14 &  Y  &  Hipp \\
G 249-37\tablenotemark{c}  & 60625 & 635007 & 5000 & -0.13 & 4.5 & 0.8 & -26 & 1 & -23 & 1 & -7 & 1 &  Y  &  CLLA94 \\
HIP 29316  & 61055 & 101905 & 4400 & -0.11 & 2 & 1.9 & 26 & 1 & -40 & 2 & -20 & 1 &  Y  &  Hipp \\
LP 720-6  & 62238 & -125305 & 4600 & -1.75 & 4.5 & 0.3 & 280 & 76 & -309 & 96 & 105 & 30 &  Y  &  Samir02 \\
LP 661-1  & 64854 & -45907 & 5200 & -1.98 & 4.5 & 0.5 & -257 & 101 & -192 & 138 & 99 & 48 &  N  &  Samir \\
HIP 33848\tablenotemark{c}  & 70136 & 65537 & 5200 & 0 & 4.5 & 0.8 & 17 & 1 & 15 & 1 & 4 & 1 &  Y  &  Hipp \\
NLTT 17752\tablenotemark{c}  & 72132 & -202041 & 5800 & -2.01 & 4 & 0.4 & 52 & 53 & -271 & 36 & -70 & 21 &  N  &  Samir02 \\
HIP 36827\tablenotemark{c}  & 72436 & -65348 & 5000 & -0.25 & 4.5 & 1.1 & 15 & 1 & 11 & 1 & -5 & 1 &  Y  &  Hipp \\
G 87-38  & 72819 & 375631 & 5800 & -0.63 & 4.5 & 0.7 & 195 & 177 & -1398 & 1462 & 354 & 369 &  Y  &  Samir \\
G 40-5\tablenotemark{c}  & 80435 & 152151 & 5200 & 0.06 & 4.5 & 0.8 & -38 & 1 & -52 & 1 & -7 & 2 &  Y  &  CLLA94 \\
G 51-7  & 82007 & 344212 & 4600 & -1.66 & 4.5 & 0.7 & -47 & 6 & -215 & 91 & 29 & 10 &  Y  &  Samir \\
HIP 42145\tablenotemark{c}  & 83528 & 414425 & 4800 & -0.25 & 4.5 & 0.6 & -67 & 1 & -74 & 3 & 8 & 1 &  Y  &  HIPP(vel) \\
G 9-13\tablenotemark{c}  & 83951 & 113122 & 5000 & -0.58 & 4.5 & 0.5 & 28 & 1 & -26 & 1 & -23 & 1 &  Y  &  CLLA94 \\
HIP 44526\tablenotemark{c}  & 90421 & -155451 & 4800 & -0.09 & 4.5 & 1 & 1 & 1 & -1 & 1 & -4 & 1 &  Y  &  HIPP(vel) \\
NLTT 20889  & 90446 & 392419 & 5800 & -2.5 & 4.5 & 1 & 58 & 14 & -331 & 176 & -57 & 23 &  N  &  Samir \\
G 41-29  & 91419 & 200136 & 4300 & -2.49 & 4.5 & 0.4 & 13 & 39 & -177 & 42 & 160 & 17 &  Y  &  Samir \\
G 116-51  & 94617 & 383918 & 4900 & -0.67 & 4.5 & 0.7 & -2 & 11 & -175 & 74 & 53 & 13 &  Y  &  Samir \\
G 43-30\tablenotemark{c}  & 101008 & 181113 & 5300 & 0.25 & 4.5 & 0.8 & 41 & 3 & -79 & 7 & -19 & 2 &  Y  &  CLLA94 \\
G 54-29  & 102854 & 270229 & 4700 & -2.05 & 4 & 0.4 & -179 & 56 & -372 & 160 & -24 & 72 &  Y  &  Samir02 \\
LP 612-37  & 111935 & -32750 & 4000 & -1.83 & 4.5 & 0.3 & 89 & 6 & -124 & 16 & 191 & 5 &  Y  &  Samir02 \\
HIP 56461\tablenotemark{c}  & 113436 & 400923 & 4800 & -1.37 & 4.5 & 0.4 & -94 & 33 & -345 & 94 & -61 & 5 &  Y  &  Samir02 \\
LP 319-42  & 115232 & 273051 & 4600 & -1.8 & 4.5 & 0.4 & -184 & 45 & -203 & 26 & 7 & 50 &  Y  &  Samir02 \\
LHS 343  & 125624 & 154143 & 3900 & -2.63 & 4 & 0.5 & 190 & 60 & -625 & 230 & 59 & 44 &  Y  &  Samir02 \\
ROSS 471  & 132041 & -122606 & 4600 & -2.13 & 4.5 & 0.5 & -88 & 37 & -97 & 43 & -15 & 6 &  Y  &  Samir02 \\
G 63-40  & 133648 & 190513 & 4600 & -1.95 & 4.5 & 0.3 & -160 & 55 & -351 & 134 & -58 & 12 &  Y  &  Samir02 \\
NLTT 35929  & 135935 & 224757 & 5200 & -0.57 & 4.5 & 0.5 & -38 & 21 & -137 & 77 & -13 & 14 &  Y  &  Samir02 \\
G 136-67  & 150816 & 150948 & 5000 & -0.5 & 4.5 & 0.6 & -88 & 22 & -65 & 18 & 43 & 18 &  Y  &  Samir02 \\
G 181-19  & 165910 & 345200 & 4800 & -1.66 & 4.5 & 0.3 & -18 & 13 & -324 & 71 & 129 & 70 &  Y  &  Samir02 \\
G 140-16  & 175248 & 143837 & 6000 & -0.32 & 3.75 & 1.4 & 842 & 754 & -835 & 689 & -471 & 390 &  N  &  Samir02 \\
LP 393-3  & 193307 & 235951 & 4300 & -1.92 & 4.5 & 0.7 & -4 & 21 & -260 & 12 & -67 & 12 &  Y  &  Samir02 \\
G 23-7  & 193544 & 14342 & 5400 & -0.68 & 4.5 & 0.6 & 33 & 20 & -170 & 26 & -24 & 8 &  Y  &  Sandage86 \\
LP 813-13  & 194343 & -155524 & 6000 & -1.42 & 4.5 & 1 & 138 & 31 & -292 & 83 & -75 & 22 &  N  &  Samir02 \\
HIP 97174  & 194503 & 291943 & 4700 & -0.83 & 4.5 & 0.3 & -105 & 6 & -57 & 3 & 12 & 2 &  Y  &  Hipp \\
NLTT 48700  & 200516 & 311210 & 4800 & -1.9 & 4.5 & 0.2 & -275 & 4 & -232 & 50 & 40 & 37 &  Y  &  Samir02 \\
G 231-39  & 211833 & 522336 & 4600 & -1.3 & 4.5 & 0.3 & 260 & 58 & -226 & 4 & 9 & 6 &  Y  &  Samir02 \\
NLTT 51106  & 212214 & -270217 & 4800 & -1.94 & 4.5 & 0.4 & -223 & 90 & 50 & 18 & -182 & 67 &  Y  &  Samir02 \\
G 126-2  & 212907 & 121059 & 4500 & -1.42 & 4.5 & 0.9 & 175 & 32 & -14 & 38 & -242 & 1 &  Y  &  Sandage86 \\
LP 873-63  & 213139 & -214812 & 5100 & -1 & 4.5 & 0.3 & 55 & 7 & -171 & 79 & -119 & 34 &  Y  &  Samir02 \\
WT 2218  & 220744 & -221453 & 6200 & -1.46 & 4.5 & 1 & -336 & 232 & -275 & 228 & -116 & 162 &  N  &  Samir02 \\
LP 759-73  & 221840 & -100812 & 5400 & -1.85 & 4.5 & 0.9 & -277 & 62 & -212 & 29 & 55 & 51 &  N  &  Samir02 \\
LP 875-62  & 222601 & -205031 & 4600 & -1.03 & 4.5 & 0.4 & 160 & 57 & -90 & 35 & 20 & 8 &  Y  &  Samir02 \\
HIP 112486  & 224704 & 295108 & 5000 & -1.5 & 4.5 & 0.5 & -2045 & 6852 & -1104 & 2808 & -1412 & 5086 &  Y  &  Samir02 \\
NLTT 55528  & 230013 & 215631 & 5800 & -2.61 & 4.5 & 1 & -548 & 351 & -307 & 77 & -38 & 89 &  N  &  Samir02 \\
G 273-76  & 233457 & -203854 & 4800 & -1.9 & 4.5 & 0.2 & 36 & 25 & -190 & 61 & -26 & 15 &  Y  &  Samir02 \\
G 157-88  & 233757 & -74542 & 5000 & -0.15 & 4.5 & 0.6 & 250 & 140 & -251 & 127 & -23 & 46 &  Y  &  Samir02 \\
G 273-105  & 234117 & -194748 & 4800 & -1.31 & 4.5 & 0.5 & 200 & 84 & -106 & 48 & 5 & 4 &  Y  &  Samir02 \\
G 266-8  & 235614 & -262559 & 5200 & -0.81 & 4.5 & 0.3 & -106 & 28 & -171 & 42 & -16 & 10 &  Y  &  Samir02 \\
G 30-31  & 235754 & 83646 & 4200 & -1.86 & 4.5 & 0.4 & 15 & 13 & -71 & 23 & -104 & 18 &  Y  &  Sandage86 \\
G 129-57\tablenotemark{c}  & 235831 & 203424 & 4700 & -0.82 & 4.5 & 0.3 & -45 & 10 & -25 & 3 & 14 & 2 &  Y  &  Sandage86 \\
\enddata 

\tablenotetext{a}{hhmmss}
\tablenotetext{b}{ddmmss}
\tablenotetext{c}{Data taken with R=60000}

\tablerefs{
(CLLA94) = \citet{carney94};
(Hipp) = \hipparcos catalog;
(NLTT) = \citet{salim02};
(Sandage86) = \citet{sandage86}
}

\end{deluxetable}

%% file: tab2.tex
\begin{deluxetable}{lcccccr} 
\tabletypesize{\tiny}
\tablecolumns{7} 
\tablewidth{0pc} 
\tablecaption{Comparison with literature\label{tab:comp}}
\tablehead{ 
\colhead{}& \multicolumn{2}{c}{This study} &
\colhead{} & \multicolumn{2}{c}{Literature} &
\colhead{}\\
\cline{2-3}  \cline{5-6}\\
\colhead{Star}& \colhead{\teff} &
\colhead{[Fe/H]}& \colhead{} &
\colhead{\teff} & \colhead{[Fe/H]} & 
\colhead{Source}
}
\startdata 
LP 644-47 & 5400 & -1.47 &  &  & -2.07 & RN91 \\
 & 5400 & -1.47 &  & 5341 & -1.63 & CLLA94 \\
LP 705-89 & 5800 & -0.97 &  &  & -1.12 & RN91 \\
G 72-18 & 4900 & -1.74 &  & 4982 & -1.86 & CLLA94 \\
G 39-1 & 5000 & -0.06 &  & 4867 & -0.61 & CLLA94 \\
HIP 19814 & 5400 & -0.71 &  & 5378 & -0.69 & CLLA94 \\
G 83-26 & 4500 & -1.57 &  &  & -2.30 & RN91 \\
HIP 24289 & 5000 & -0.66 &  & 5032 & -0.77 & CLLA94 \\
G 249-37 & 5000 & -0.13 &  & 4954 & -0.27 & CLLA94 \\
LP 661-1 & 5200 & -1.98 &  &  & -2.03 & RN91 \\
G 40-5 & 5200 & 0.06 &  & 5012 & -0.07 & CLLA94 \\
G 9-13 & 5000 & -0.58 &  & 5049 & -0.27 & CLLA94 \\
G 43-30 & 5300 & 0.25 &  & 5070 & -0.21 & CLLA94 \\
G 181-19 & 4800 & -1.66 &  & 4892 & -1.56 & CLLA94 \\
LP 813-13 & 6000 & -1.42 &  &  & -2.10 & RN91 \\
LP 873-63 & 5100 & -1.00 &  &  & -0.83 & RN91 \\
LP 759-73 & 5400 & -1.85 &  &  & -1.63 & RN91 \\
G 273-105 & 4800 & -1.31 &  &  & -2.17 & RN91 \\
\enddata 

\tablerefs{
(CLLA94) = \citet{carney94};
(RN91) = \citet{ryan91};
}

\end{deluxetable}